\documentclass{ws-procs975x65}
\newcommand{\R}{\mathbb R}
\def\lg{\langle }
\def\rg{\rangle }

\def\lu{\mbox{\large 1}}
\def\ud{\mathrm{d}}

\def\mq{\mathfrak{q}}
\def\mp{\mathfrak{p}}

\begin{document}
\title{Affine Coherent States in Quantum Cosmology}

\author{Ma\l kiewicz, Przemys\l aw}

\address{$^1$APC, Univ Paris Diderot, CNRS/IN2P3, CEA/Irfu, Obs de Paris, Sorbonne Paris Cit\'e, France\\
$^2$National Centre for Nuclear Research, Ho\.za 69, Warsaw, Poland\\
E-mail: przemyslaw.malkiewicz@ncbj.gov.pl}

\begin{abstract}
A brief summary of the application of coherent states in the examination of quantum dynamics of cosmological models is given. We discuss quantization maps, phase space probability distributions and semiclassical phase spaces. The implementation of coherent states based on the affine group resolves the hardest singularities, renders self-adjoint Hamiltonians without boundary conditions and provides a completely consistent semiclassical description of the involved quantum dynamics. We consider three examples: the closed Friedmann model, the anisotropic Bianchi Type I model and the deep quantum domain of the Bianchi Type IX model.
\end{abstract}

\keywords{quantum cosmology; big bounce; semiclassical cosmology; affine quantization; coherent states}

\bodymatter

\section{Introduction}

The Hamiltonian constraint for the spatially homogenous and diagonal Bianchi class A models, discussed e.g. in Ref.~\refcite{hom}, and filled with a perfect fluid reads:
$$
\mathsf{H}_{tot}=\mathsf{H}_g+\mathsf{H}_f=\frac{Ne^{-3\Omega}}{24}\left(-p_{\Omega}^2+p_+^2+p_-^2+24e^{4\Omega}V(\beta_{\pm})+24e^{3(1-w)\Omega}p_T\right)\approx 0
$$
where $p_T>0$ is the momentum of barotropic fluid with $p=w\rho$. The Misner variables $\Omega$ and $\beta_{\pm}$ describe respectively the isotropic and anisotropic components of the evolving three-geometry, whose Ricci curvature is encoded in the potential $V$. We solve the above constraint by setting $T$ as a clock and removing $p_T$ from the framework. We find the true non-vanishing Hamiltonian:
$$
\mathsf{H}=\frac{1}{24}\left(p^2-c_1\frac{p_+^2+p_-^2}{q^2}-c_2q^{\frac{2w+\frac{2}{3}}{w-1}}V(\beta_{\pm})\right),~~q=\frac{e^{\frac{3}{2}(1-w)\Omega}}{\frac{3}{2}(1-w)},~~p=p_{\Omega}e^{-\frac{3}{2}(1-w)\Omega},
$$
which generates dynamics in the physical phase space $(q,p,\beta_{\pm},p_{\pm})\in\mathbb{R}_+\times\mathbb{R}^3$. Note that $q>0$ is positive. The dynamics resembles the motion of a particle in the half-line $q>0$ in some potential. The potential is due to the shear and the intrinsic curvature. The particle hits the singularity $q=0$ at a finite value of clock $T$. 

The quantization of anisotropic pairs $(\beta_{\pm},p_{\pm})$, which are real, can be based on the usual canonical prescription. On the other hand, the isotropic variables $(q,p)$ have the range of the half-plane and their quantization must follow a suitably adapted prescription. In fact, the cosmological phase space $(q,p)$ may be identified with the affine group, which is given by the multiplication law: $$(q',p')\cdot (q,p) = (q'q,\frac{p}{q'}+p').$$ Its unitary irreducible and integrable representation in $\mathcal{H}={L}^2(\R^{\ast}_+, \ud x)$ reads: $$
U(q,p) \psi(x) = e^{i\mp x } \frac{1}{\sqrt{\mq}} \psi(x/\mq)
.$$  Given a \underline{normalized} vector $\psi_0 \in \mathcal{H}$, a continuous family of unit vectors may be defined 
$$
|q,p\rg = U(q,p)|\psi_0 \rg\, , \quad \lg x | q,p\rg = e^{i\mp x}\frac{1}{\sqrt{\mq}} \psi_0(x/\mq)\,. 
$$
By the virtue of Schur's lemma they resolve unity:
$$
\int \frac{\ud q \ud p}{2\pi} | q,p\rg\lg q,p| = c_{-1}\cdot\lu,~~~c_{-1} <\infty.
$$
The following quantization map respects the symmetries of the cosmological phase space:
$$
f(q,p)\mapsto A_{f}:=\int \frac{\ud q \ud p}{2\pi c_{-1}} f(q,p) | q,p\rg\lg q,p| .
$$
The above quantization is covariant with respect to the affine group rather then Weyl-Heisenberg group. Moreover, it has all the demanded properties: (i) it is linear, (ii) to $f(q,p)\equiv 1$ it assigns identity and (iii) to semi-bounded $f(q,p)$ it assigns semi-bounded operators. Since there exist infinitely many possible families of states $| q,p\rg$, which depend on the choice of $\psi_0$, there follow infinitely many quantization maps. Such a parameter-dependent quantization procedure seems well-suited for investigating possible singularity resolutions in quantum gravity.

The affine quantization of canonical coordinates $q$ and $p$ was studied in Ref.~\refcite{FRW} and reads:
$$q \mapsto A_q = \frac{c_0}{c_{-1}} Q\, ,~~ p \mapsto A_p = P\, ,$$
where $Q$ and $P$ are the position and momentum operators defined on the half-line, $x>0$ and the constants $c_{\alpha}$ depend on the fiducial vector $\psi_0$. Note that $P$ is not self-adjoint on the half-line. The choice of  $\psi_0$ can be constrained in a way to obtain the canonical commutation rule for the basic variables, $[A_q,A_p]=1$, i.e. $c_0=c_{-1}$. 

\section{Basic example}
\begin{figure}[t!]
\centering
\parbox{1.21in}{\includegraphics[width=1.2in]{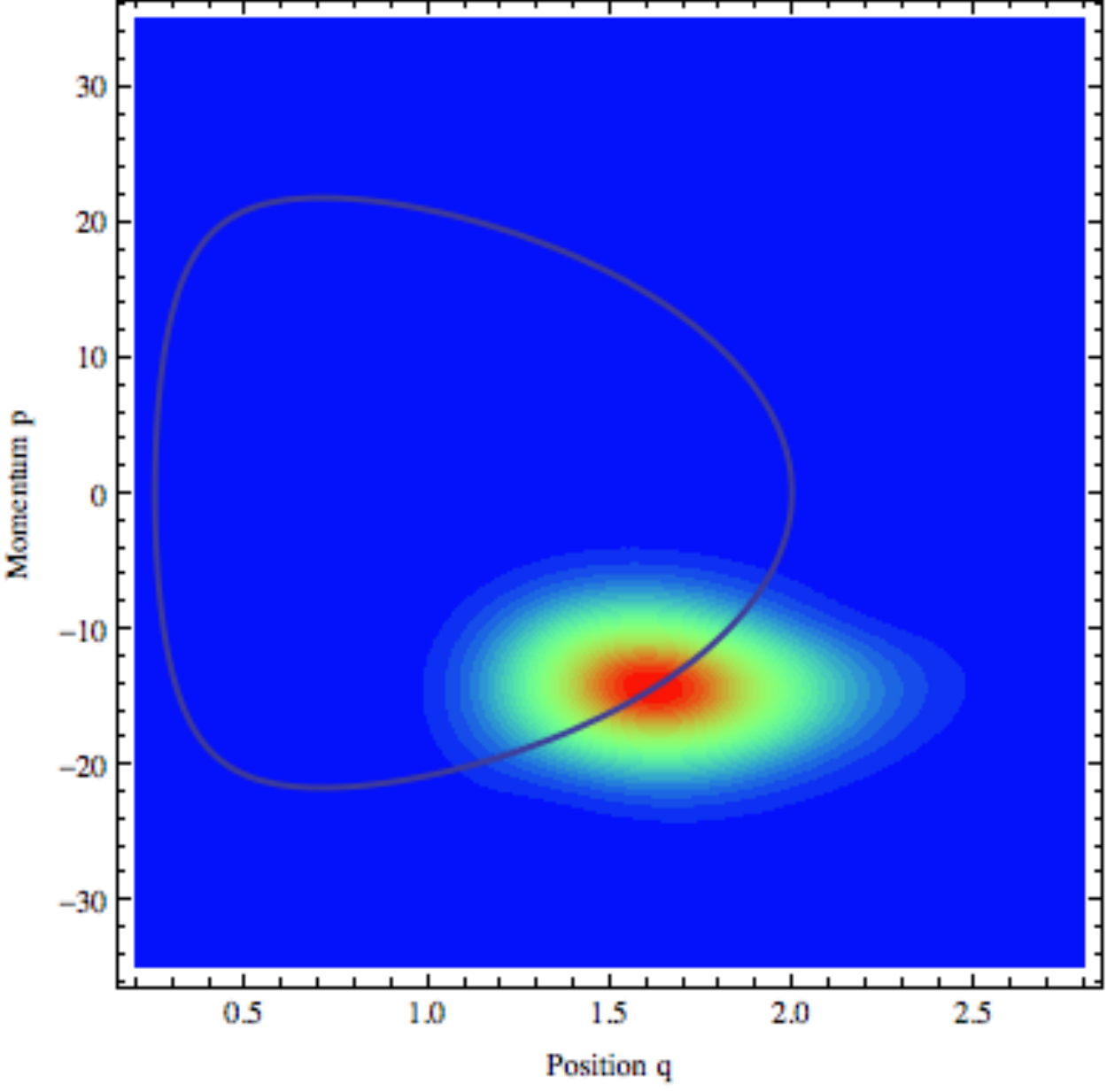}}
\parbox{1.21in}{\includegraphics[width=1.2in]{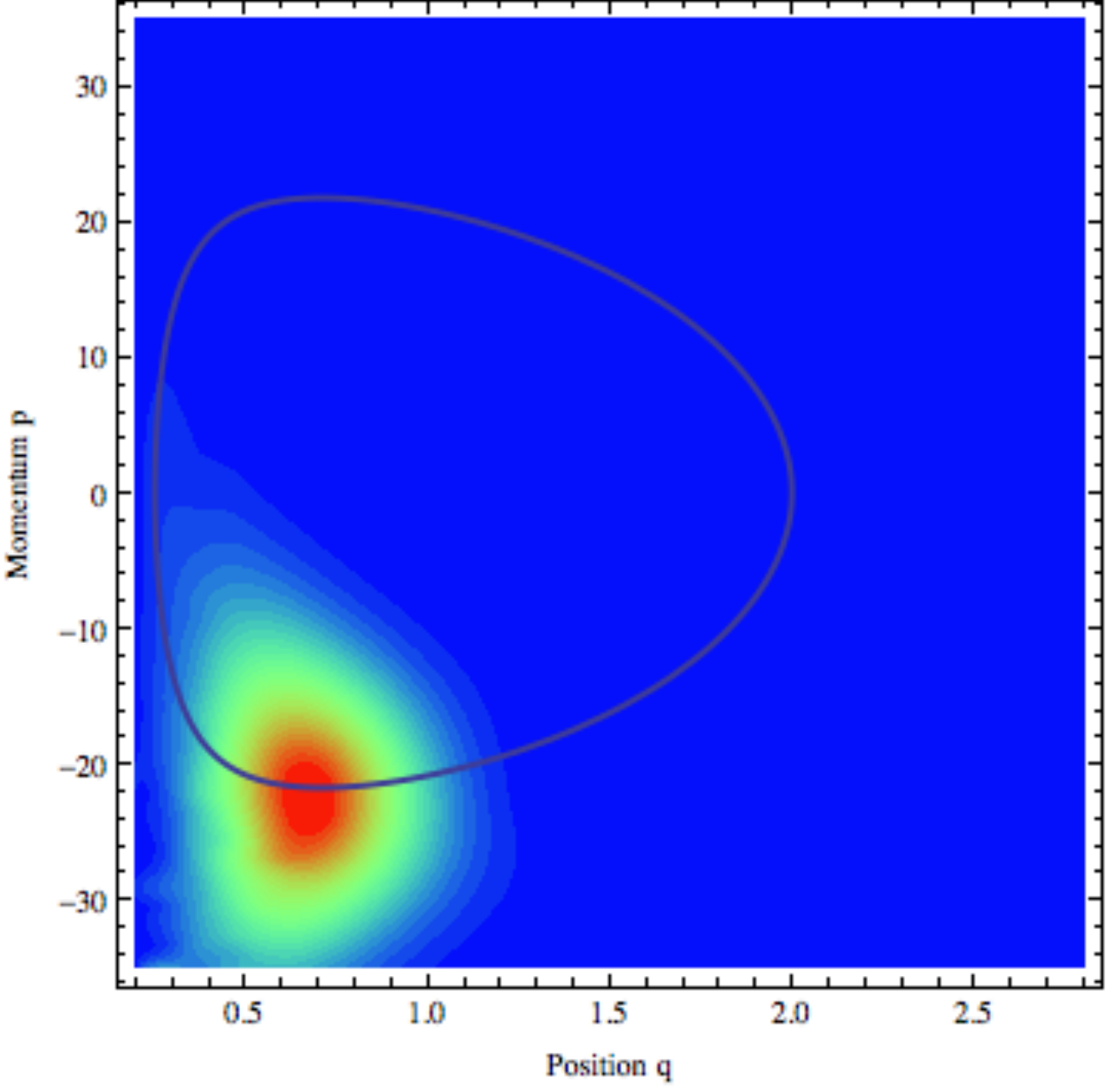}}
\parbox{1.21in}{\includegraphics[width=1.2in]{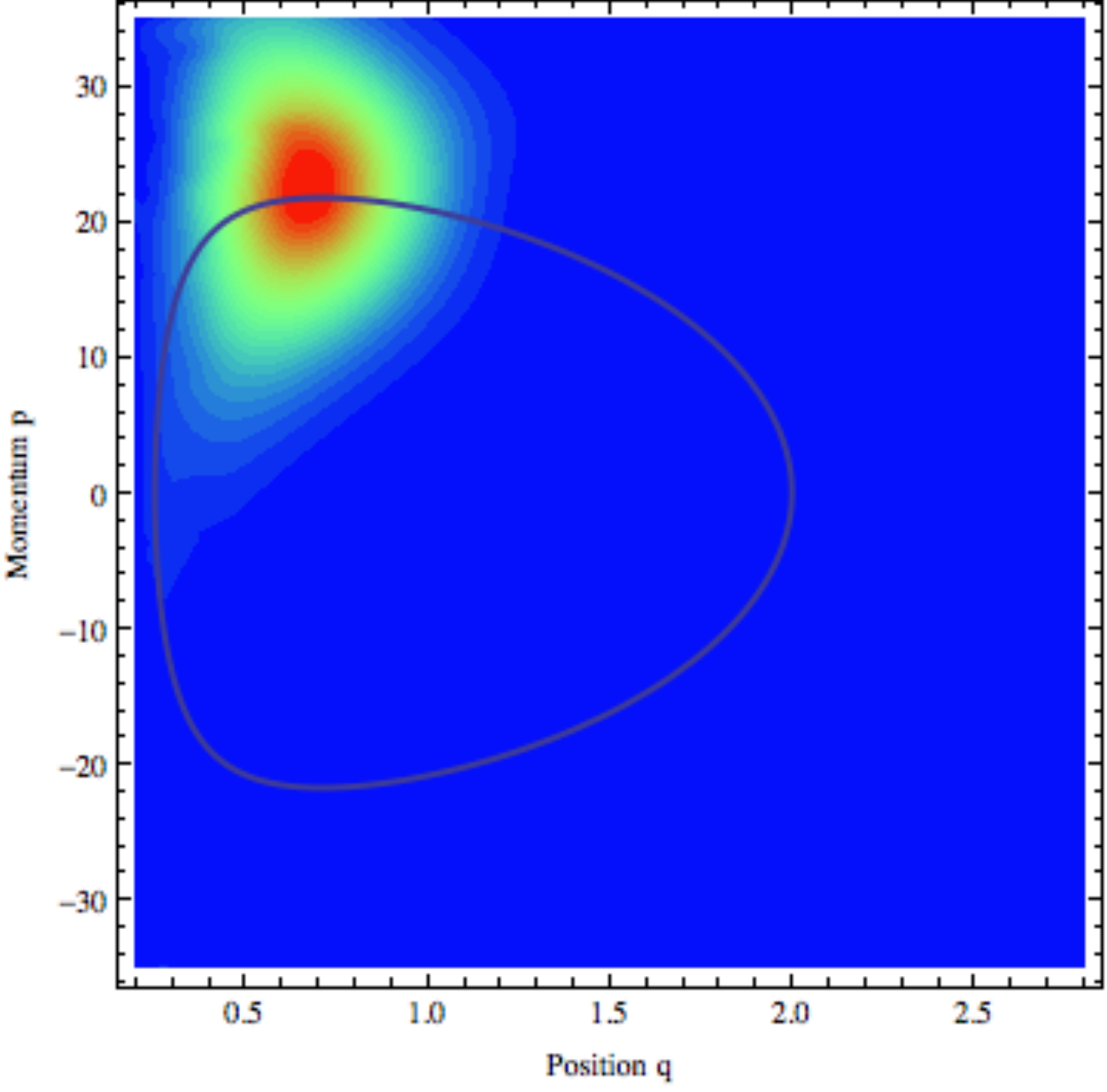}}
\parbox{1.21in}{\includegraphics[width=1.2in]{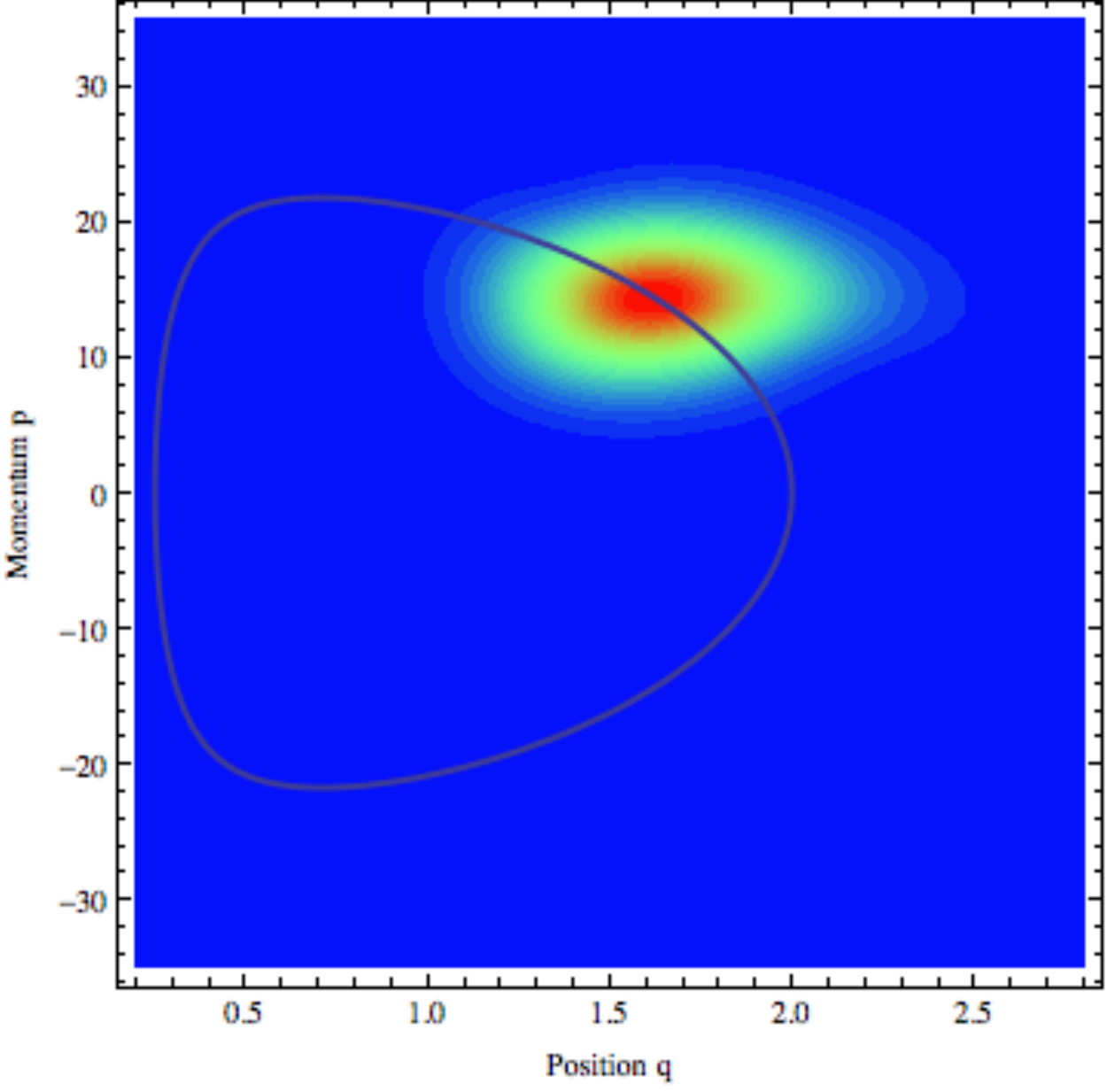}}
\caption{Phase space distributions $\rho_{q_0,p_0}(p,q,T)$ at different times equally spaced (from left to right). The ranges in $q$ and $p$ are respectively $[0.2, 2.8]$ and $[-35, +35]$. Increasing values of the functions are encoded by the colors from blue to red.} 
\label{figure1}
\end{figure}

Let us start with an analysis of the closed Friedmann model. For brevity, we set radiation as the content of the universe, $w=\frac{1}{3}$. Then the true Hamiltonian reads:
$$
\mathsf{H}=\frac{1}{24}p^2+k q^2~,
$$
where $k=1$. The quantization of the kinetic term reads (cf. Ref.~\refcite{FRW}):
$$p^2 \mapsto A_{p^2} = P^2 + \frac{K(\psi_0)}{Q^2}+\frac{J(\psi_0)}{2}(PQ^{-1}+Q^{-1}P)\, .  $$
The constants $K$ and $J$ depend on the fiducial vector $\psi_0$. $J$ is non-vanishing only for complex fiducial vectors and will be disregarded in what follows. We note that $A_{p^2}$ is an ordinary differential Sturm-Liouville operator, singular at the end point $x=0$. It follows that $A_{p^2}$, if defined on the domain of smooth compactly supported functions, is essentially self-adjoint for $K \ge 3/4$, whereas for $K<3/4$ the deficiency indices of $A_{p^2}$ are $(1, 1)$ and thus more self-adjoint extensions exist requiring setting the boundary condition. We choose the fiducial vector $\psi_0$ in such a way that $K \ge 3/4$ in order to comply with essential self-adjointness of the Hamiltonian.

Given a state $|\phi\rg$, coherent states define the so-called phase space representation $\Phi$ and the associated phase space probability distribution $\rho_{\phi}$:
$$
\Phi(q,p):=\frac{1}{\sqrt{2\pi c_{-1}}}\lg q,p |\phi\rg,~~\rho_{\phi}(q,p):=\big|\Phi(q,p)\big|^2=\frac{1}{2\pi c_{-1}}|\lg q,p |\phi\rg|^2~.
$$
Having a well-defined quantum Hamiltonian, we can study the evolution of the phase space probability distribution:
$$
\rho_{\phi}(q,p,T)=\frac{1}{2\pi c_{-1}}|\lg q,p |e^{-iA_{h}T}|\phi\rg|^2~.
$$
In  Fig.\;\ref{figure1} we plot the quantum evolution of the phase space probability distribution of the Friedmann universe by setting the coherent state $|\phi\rg=| 2,0 \rg$ as the initial state. See Ref. \refcite{FRW}.

\begin{figure}[t!]
\centering
\parbox{1.81in}{\includegraphics[width=1.8in]{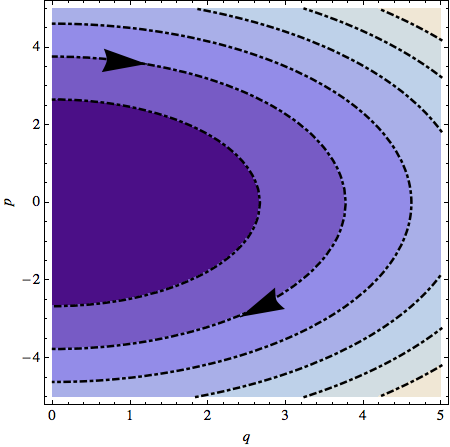}}\hspace*{25pt}
\parbox{1.81in}{\includegraphics[width=1.8in]{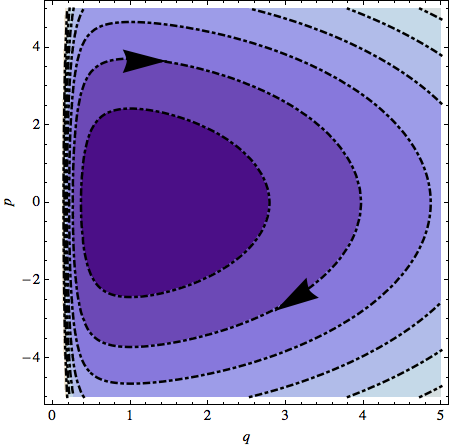}}
\caption{Contour plots of the classical and semiclassical Hamiltonians for the closed Friedmann universe. The singularity $q=0$ on the left is replaced with a bounce on the right.}
\label{2a}
\end{figure}

Furthermore, the coherent states are used to map quantum operators back to the classical world, where they become semiclassical observables replacing the classical ones in the phase space. They are called `lower symbols' and the lower symbol of $A_f$ is defined as:
$$
\check{f}(q,p):=\lg q,p |A_f | q,p\rg=\int\frac{\ud q'\ud p'}{2\pi c_{-1}}|\lg q',p'|q,p\rg |^2f(q',p')
$$
We require that $\check{q}=q$ and $\check{p}=p$. The lower symbol of Hamiltonian is shown (cf. Ref.~\refcite{klauderscm}) to generate semiclassical dynamics. The contour-plots of the classical and semiclassical Hamiltonians for the Friedmann model are presented in Fig. \ref{2a}.

\section{Non-oscillatory anisotropic singularities}

\begin{figure}[t!]
\centering
\parbox{1.81in}{\includegraphics[width=1.8in]{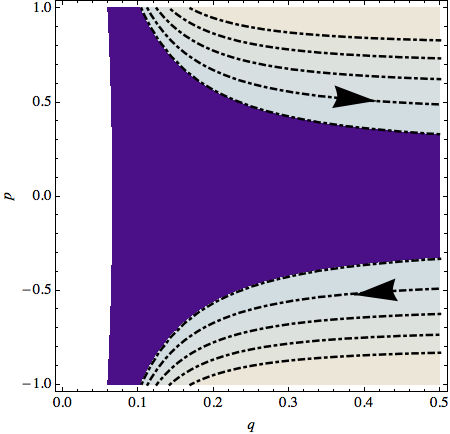}}
\hspace*{25pt}
\parbox{1.81in}{\includegraphics[width=1.8in]{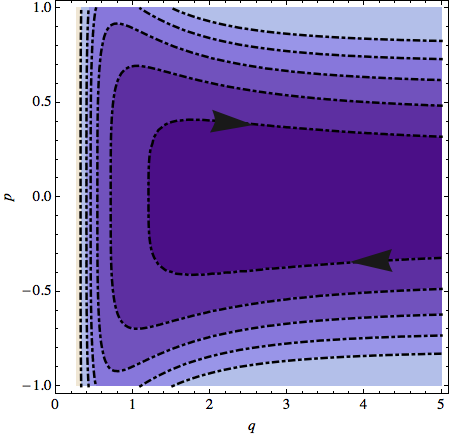}}
\caption{Contour plots of the classical and semiclassical Hamiltonian for Bianchi Type I universe. The separating region $\mathsf{H}<0$ is visible on the left.}
\label{2b}
\end{figure}

The strongest singularity is exhibited by the anisotropic Bianchi I model. The non-vanishing shear in this model fuels the gravitational collapse to such an extent as to produce diverging trajectories in the phase space $(q,p)$. The physical Hamiltonian reads:
$$
\mathsf{H}=\frac{1}{24}\left(p^2-c_1\frac{p_+^2+p_-^2}{q^2}\right)
$$
Due to the positive energy density of the fluid, the physical Hamiltonian must satisfy $\mathsf{H}>0$, which in the presence of the shear is non-trivial. The cosmological phase space $(q,p)$ consists in expanding and collapsing trajectories, which originate and terminate respectively in the singularity $q=0$ and which are separated from each other by the non-physical region, $\mathsf{H}<0$. Affine quantization naturally smooths (quantum effect) the positivity constraint so the contracting quantum states become dynamically connected with the expanding ones. Specifically, we implement the positivity constraint in the quantum model by quantizing $\theta(\mathsf{H})\mathsf{H}$ instead of $\mathsf{H}$, where $\theta(\cdot)$ is the Heaviside function. We have obtained the respective quantum dynamics generator, $A_{\theta(\mathsf{H})\mathsf{H}}$, in Ref.~\refcite{B1}. The lower symbol, which encodes semiclassical features of the quantized dynamics, can be computed numerically and its approximation can be given in terms of analytic functions of $q,~p,~k$ and $\nu$ (see Ref.~\refcite{B1}). The classical and the corresponding semiclassical Hamiltonians are contour-plotted in Fig. \ref{2b}. We notice that due to quantization the classically forbidden region $\mathsf{H}<0$ becomes semi-classically accessible. Contracting universes close enough to the singularity bend and extend across this region to emerge smoothly in the expanding branch. The singularity is resolved with a bounce. On the contrary, it is intuitively understood that the Wheeler-DeWitt equation can not be successful in dealings with this sort of hard, shear-fuelled singularities.

\section{Oscillatory singularities}

\begin{figure}[t!]
\centering
\parbox{1.81in}{\includegraphics[width=1.8in]{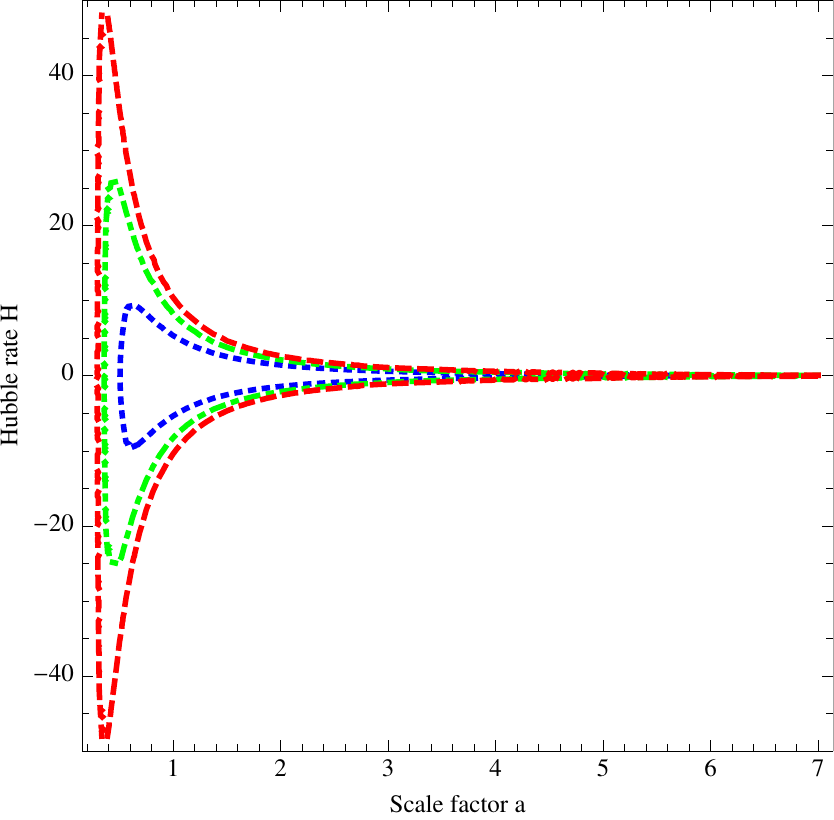}}
\caption{Adiabatic non-singular evolution of the quantum Mixmaster universe.}
\label{2c}
\end{figure}

The generic local behavior of cosmological spacetimes on their approach to the singularity becomes asymptotically, by virtue of the BKL conjecture, identical with the dynamics of Bianchi VIII/IX models, in which all familiar forms of matter become dynamically negligible. In what follows we study the Hamiltonian 
constraint of the Mixmaster universe (the vacuum Bianchi Type IX model):
$$
\mathsf{H}_{tot}= \frac{9}{4} \, p^2+36^2q^{2/3} -\mathsf{H}_q\approx 0 \,,~~\mathsf{H}_q :=\frac{ p_+^2+p_-^2}{q^2}+36 q^{2/3} V(\beta)$$
The first part of $\mathsf{H}$ describes the energy of the isotropic expansion and the isotropic curvature potential energy, whereas $\mathsf{H}_q$ describes the energy of the anisotropic oscillations of the three-geometry. When the universe approaches the singularity, anisotropic oscillations become very fast with respect to the contraction rate. Therefore, in quantum theory, in analogy with molecular physics, one can assume a fixed eigenstate of quantized $\mathsf{H}_q$, and consider the dynamics of $q$ in the potential induced by this eigenstate. This is the so-called Born-Oppenheimer approximation. Note that the energy of oscillations is dynamically significant and can not be neglected. For low eigenstates, it reads  $E_N(q) \simeq \frac{24 \hbar}{q^{2/3}} \sqrt{2 \frak{K}_2
\frak{K}_3} \, (N+1)$. The semiclassical constraint was derived in Ref. \refcite{B9} and reads:
$$
\check{\mathsf{H}}_{tot}(q,p,N)  =
\frac{9}{4} \left( p^2 +
\frac{\hbar^2 \frak{K}_4}{q^2} \right) +36
\frak{K}_5 q^{2/3} - \frac{24 \hbar}{q^{2/3}} \frak{K}_6 (N+1)\approx 0
$$
where $\frak{K}_i$ are constant dependent on $\psi_0$. The corresponding Friedmann-like equation is plotted in Fig. \ref{2c}. The singularity is resolved with a bounce. 

It was shown that when the BO approximation breaks down, then a huge production of anisotropy may take place at the bounce. In order to maintain the proper balance between the anisotropic and isotropic expansion energies, this must be accompanied by a vigorous inflationary-like phase of accelerated isotropic expansion. More details may be found in Ref.~\refcite{IB9}.

\section{Conclusion}
Coherent states prove to be very useful to deal with the singularity problem in the spatially homogenous models. They are used to define a covariant quantization (respecting the symmetries of the cosmological phase space), which leads to the resolution of singularities. They provide a fully consistent semiclassical description. Presently, in Ref.~\refcite{B9VA}, we use coherent states to develop the so-called vibronic framework for the quantum theory of the oscillatory models, which extends beyond the regime of validity of the BO approximation.

\section*{Acknowledgements}
Author was supported by the ``Mobilno\'s\'c Plus" fellowship from Ministerstwo Nauki i Szkolnictwa Wy\.zszego.

\end{document}